\documentclass[preprint]{aastex631}
\usepackage{natbib}
\usepackage{graphicx}
\usepackage[utf8]{inputenc}
\usepackage[english]{babel}

\begin{document}

\title{Panchromatic HST/WFC3 Imaging Studies of Young, Rapidly Evolving Planetary Nebulae. II. NGC 7027}
\author{Paula Moraga Baez}
\affiliation{School of Physics and Astronomy and Laboratory for Multiwavelength Astrophysics, Rochester Institute of Technology, Rochester, NY 14623, USA}
\author{Joel H. Kastner}
\affiliation{Chester F. Carlson Center for Imaging Science and Laboratory for Multiwavelength Astrophysics, Rochester Institute of Technology, 54 Lomb Memorial Drive, Rochester, NY 14623, USA}
\author{Bruce Balick}
\affiliation{Department of Astronomy, University of Washington, Seattle, WA 98195, USA}
\author{Rodolfo Montez, Jr.}
\affiliation{Harvard-Smithsonian Center for Astrophysics, 60 Garden Street, Cambridge, MA 02138, USA}
\author{Jesse Bublitz}
\affiliation{Green Bank Observatory, 155 Observatory Road, Green Bank, WV 24944, USA}

\begin{abstract}
The iconic planetary nebula (PN) NGC 7027 is bright, nearby ($D \sim1$ kpc), highly ionized, intricately structured, and well observed. This nebula is hence an ideal case study for understanding PN shaping and evolution processes. Accordingly, we have conducted a comprehensive imaging survey of NGC 7027 comprised of twelve HST Wide Field Camera 3 images in narrow-band and continuum filters spanning the wavelength range 0.243--1.67 $\mu$m. The resulting panchromatic image suite reveals the spatial distributions of emission lines covering low-ionization species such as singly ionized Fe, N, and Si, through H recombination lines, to more highly ionized O and Ne. These images, combined with available X-ray and radio data, provide the most extensive view of the structure of NGC 7027 obtained to date. Among other findings, we have traced the ionization structure and dust extinction within the nebula in sub-arcsecond detail; uncovered multipolar structures actively driven by collimated winds that protrude through and beyond the PN's bright inner core; compared the ionization patterns in the WFC3 images to X-ray and radio images of its interior hot gas and to its molecular outflows; pinpointed the loci of thin, shocked interfaces deep inside the nebula; and more precisely characterized the central star. We use these results to describe the recent history of this young and rapidly evolving PN in terms of a series of shaping events. This evolutionary sequence involves both thermal and ram pressures, and is far more complex than predicted by extant models of UV photoionization or winds from a single central progenitor star, thereby highlighting the likely influence of an unseen binary companion.
\end{abstract}


\section{Introduction} \label{sec:intro}

A planetary nebula (PN) is generated when an intermediate mass star (~0.8-8 M$_\odot$) evolves off the asymptotic giant branch (AGB), having ejected its dusty AGB envelope at mass loss rates exceeding $\sim10^{-5}$ $M_\odot$ yr$^{-1}$ \citep[][]{Hofner2018}. Within a few tens or hundreds of years after AGB envelope ejection, depending on the star's initial mass \citep[e.g.,][]{Miller2016}, the increasing wind speed of the rapidly transitioning, mass-losing star --- from $\sim$10 km s$^{-1}$ on the AGB to $\sim$1000 km s$^{-1}$, post-AGB --- sweeps the AGB material into a dense shell, even as the UV from the newly exposed, hot ($T_{eff} \sim$ 200 kK) AGB core begins to ionize the AGB ejecta, forming the PN. The ``classical'' result of the foregoing post-AGB evolutionary processes is a spherically symmetric, thick-walled PN with semi-evacuated interior that is bright in atomic emission lines across the UV through IR wavelength range \citep{Kwok1978,Sch2018,Schon2005,Toala2014}.

However, a significant fraction of PNe display profoundly axisymmetric and/or point-symmetric structures \citep[see, e.g.,][and references therein]{Parker2006}. Furthermore, it has been known for decades that many of these highly non-spherical objects harbor large masses of residual molecular gas and dust that are directly descended from AGB envelope material \citep[e.g.][]{Kastner1996,Bublitz2019}. The present consensus is that the non-spherical shapes and large residual molecular masses of these PNe are the direct consequence of the presence and influence of binary companions during AGB evolution \citep[e.g.,][and references therein]{DeMarco2017}. Binary companions can influence mass-launching and collimation mechanisms, which then shape PNe by way of ram pressure in collimated flows or jets \cite[e.g.,][]{Balick2019}. The details of these binary-induced PN shaping processes remain elusive, however; fast mass loss driven by a binary companion’s accretion disk as well as common envelope evolution are leading hypotheses to explain the origin of collimated outflows \citep[e.g.,][]{Garcia2018,Zou2020,Zou2022}. 

Hubble Space Telescope (HST) emission-line imaging studies of the most recently formed, and hence most rapidly evolving, PNe provide particularly effective means to understand PN ionization and shaping processes \citep[e.g.,][]{SahaiTrauger1998}. 
In HST's Cycle 27, NGC 7027 was one of the first two PNe to be targeted for a comprehensive, contemporaneous set of emission-line images, from near-UV through optical to near-IR, with the Wide Field Camera 3 (WFC3) \citep{Kastner2020}. We selected NGC 7027 for such a study because it is among the youngest and most rapidly evolving PN within $\sim$1 kpc of the Sun \citep[$D = 0.89$ kpc;][]{Masson1989}. Indeed, NGC 7027 represents an ``industry standard'' for emission-line searches and surveys, thanks to the sheer number of bright lines it displays across the electromagnetic spectrum, from radio to X-ray \citep[e.g.,][]{Zhang2005,Zhang2008,Montez2018,Neufeld2020}. 

Previous HST imaging using its Wide Field Planetary Camera 2 (WFPC2) and Near-Infrared Camera and Multi-object Spectrometer (NICMOS) have established the remarkably complex structure of the nebula \citep[e.g.,][and references therein]{Latter2000,Sch2018,Guerrero2020}. Its distinguishing features include a bright, elliptical inner region, a dusty equatorial belt, and multiple, point-symmetric collimated outflows, with the latter confirmed kinematically via ground-based near-infrared and radio interferometric imaging spectroscopy \citep[e.g.,][]{Cox2002,Nakashima2010}. Concentric ring-like or spiral structures are observed to surround the core region of NGC 7027 in HST/WFCP2 H$\alpha$ and V-band images \citep[][and references therein]{Guerrero2020}. Radio-regime proper motion analyses of the inner elliptical shell indicate a kinematic age of $\sim$600 yrs \citep{Masson1989,Z2008}, while nebular expansion analyses that are based on multi-epoch HST imaging of the extended ring system suggest a dynamical age in the range 1080--1620 yrs \citep{Sch2018,Guerrero2020}. Consistent with such a recent departure from the AGB, the nebula harbors an exceedingly hot ($\sim$200 kK) proto-white dwarf (pWD) at its center \citep{Latter2000}. 

First results from our Cycle 27 HST/WFC3 imaging programs targeting NGC 7027 and a second, well-studied PN --- the similarly young (expansion age $\sim$2000 yr), nearby ($D \sim 1.0$ kpc) bipolar PN NGC 6302 --- were presented in \citet{Kastner2020}. In \citet{Kastner2022}, we presented a detailed overview of the HST/WFC3 imaging survey of NGC 6302. Here, we present the full suite of Cycle 27 HST/WFC3 images of NGC 7027, and we highlight compelling results obtained from the image suite. In \S~2, we describe the observations and image processing that was performed to obtain the final suite of a dozen (mostly narrow-band) WFC3 images, which are presented and described in \S~3. In \S~4, we present key line ratio images that reveal the extinction and ionization structure of the nebula. In \S~5, the focus is shifted onto the central star (CS) where its extinction, luminosity, and present-day mass are evaluated. In \S~6, we develop and discuss extinction, ionization, and other lines of evidence for the presence and structure of nebula-shaping shocks. The implications of the results for the ongoing structural evolution of NGC 7027 are described in \S~7. A summary of the results of our HST/WFC3 imaging study of NGC 7027 is presented in \S~8.

\section{Observations} \label{sec:data}

\setlength{\parindent}{4ex}
Images of NGC 7027 presented in this paper were obtained with both the UVIS and NIR units of HST's WFC3 during Cycle 27 in 2019 September. The UVIS (CCD sensor) channel provides a field of view of ~2.7$'$ x 2.7$'$ and pixel scale of 0.04$''$, and the NIR channel uses a HgCdTe array with field of view of 2.27$'$ x 2.27$'$ and pixel scale of 0.13$''$. Over the range of wavelengths imaged here, the point-spread function (PSF) varies from $\sim0.025''$ (near-UV) to $\sim0.16''$ (NIR). Table \ref{tbl:summary} summarizes the WFC3 filters used to obtain images of NGC 7027 as well as the targeted emission line, the date the image was obtained, and the total exposure time of imaging through each filter. The UVIS images were obtained in 2-point GAP-LINE dither mode (DITHER-LINE for quad filter FQ243N), and the IR images were obtained in 2-point DITHERBLOB dither mode. UVIS and IR images were obtained at a position angle of --16.9$^\circ$ and --17.2$^\circ$ measured east of north, respectively.

Processing of images is described in the comprehensive HST/WFC3 imaging targeting NGC 6302 \citep{Kastner2022} and is briefly described here. The standard pipeline calibration and processing, using CALWF3 v3.5.0, included bias correction, dark current subtraction, flat field and shutter shading corrections, cosmic-ray rejection, and (for UVIS images) CTE correction. Additional post-pipeline processing was performed using the DrizzlePac package software\footnote{https://drizzlepac.readthedocs.io/en/latest/index.html}, particularly the \texttt{tweakreg} and \texttt{astrodrizzle} modules. Fine image registration was accomplished by defining the World Coordinate System of the F160W filter image via the \texttt{tweakreg} module, using Gaia Data Release 2 catalog stars as positional references. The other WFC3 images were then aligned (using \texttt{tweakreg}) with the F160W image serving as the reference. This positional calibration also included geometric distortion corrections and additional cosmic-ray corrections. The dithered exposures of each long exposure image were then merged using the \texttt{astrodrizzle} module. This process was implemented to correct for any misalignment that remained after the initial pass through HST's image reduction pipeline. We estimate that, following this image registration procedure, the WFC3 images of NGC 7027 are aligned to a relative accuracy of $\le$0.05$''$.

Evidence for pixel saturation is present in the long-exposure F502N and F656N images at the regions of highest surface brightness within the nebula, specifically $\sim$10$''$ to the northwest of the central star. Pixels that exceeded the saturation threshold of 65,000 counts (see WFC3 Instrument Handbook; \citealt{wfc3ih}) were replaced with count rates from short-exposure image pixels. This ``pixel grafting'' procedure was confined to displacements of 0$''$ to $-$10$''$ (RA) and 0$''$ to 10$''$ (dec) with respect to the position of NGC 7027's central star.

\renewcommand{\baselinestretch}{1.5}
\begin{table}
\begin{center}
\caption{\sc HST/WFC3 Imaging Survey of NGC 7027:
  Observation Summary}
\label{tbl:summary}
\footnotesize
\begin{tabular}{ccccc}
\hline
Filter & $\lambda_0$ ($\Delta \lambda$)$^a$ &  Line Targeted & Date & Exp.\ \\
         & (nm) & & & (s) \\
\hline
\hline
FQ243N & 246.8 (3.6) & [Ne {\sc iv}] $\lambda 2425$ & 2019-09-30 & 1140 \\
F343N  & 343.5 (25.0) & [Ne {\sc v}] $\lambda 3426^b$ & 2019-09-30 & 1140 \\
F487N & 487.1 (6.0) & H$\beta$ $\lambda 4861^c$ & 2019-09-30 & 1130 \\
F502N &  501.0 (6.5) & [O {\sc iii}] $\lambda\lambda 4959, 5007$ & 2019-09-30 & 1000, 30$^d$\\
F656N &  656.1 (1.8) & H$\alpha$ $\lambda 6563^c$ & 2019-09-30 & 1130, 30$^d$\\
F673N &  676.6 (11.8) & [S {\sc ii}] $\lambda\lambda 6716, 6730$ & 2019-09-30 & 1260 \\
F110W & 1153.4 (443.0) & ``YJ band'' & 2019-09-30 & 556 \\
F128N & 1283.2 (15.9) & Pa$\beta$ 1.28 $\mu$m & 2019-09-30 & 506 \\
F130N & 1300.6 (15.6) & Pa$\beta$ continuum & 2019-09-30 & 506\\
F160W & 1536.9 (268.3) & ``H band'' & 2019-09-30 & 456 \\
F164N & 1640.4 (20.9) & [Fe {\sc ii}] 1.64 $\mu$m & 2019-09-30 & 1306 \\
F167N & 1664.2 (21.0) & [Fe {\sc ii}] continuum & 2019-09-30 & 1306 \\
\hline
\end{tabular}\label{tab:filters}
\end{center}
\renewcommand{\baselinestretch}{1.0}
\tablecomments{
a) Filter pivot wavelength and effective bandwidth. b) Potential contamination from  O {\sc iii} $\lambda 3312$, $\lambda 3341$, and $\lambda 3444$  \citep{Zhang2005}. c) Potential contamination from  He {\sc ii} $\lambda 4859$ and $\lambda 6560$, respectively \citep{Zhang2005}. d) Short exposure images used for pixel grafting of saturated pixels in the long exposures.}
\end{table}

\section{The HST/WFC3 Image Suite}\label{sec:results}

\begin{figure}
    \centering
    \includegraphics[width=0.90\textwidth]{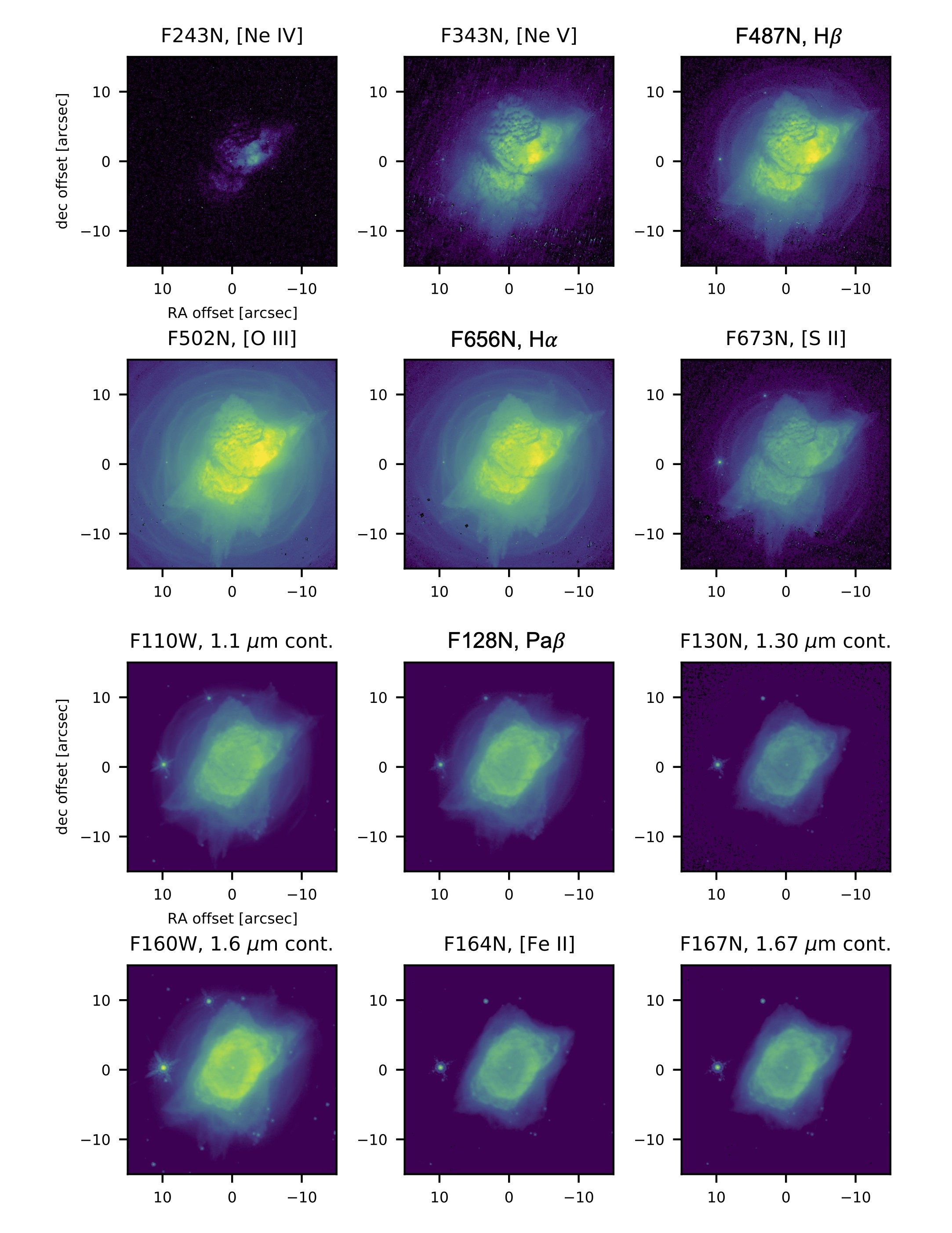}
    \caption{The complete suite of HST/WFC3 images obtained for NGC 7027. The images are oriented with N up and E to the left. The field of view in each image is 30" x 30".}
    \label{fig:rawImgs}
\end{figure}

Fig. \ref{fig:rawImgs} presents the complete suite of HST/WFC3 images obtained for NGC 7027, spanning a range from near-UV ($\sim$245 nm) to near-IR ($\sim$1665 nm). These images highlight the overall dramatic variation in the morphology of NGC 7027 over this wavelength range. Much of these differences can be attributed to the varying effects of extinction, but also reflect the intensity of the specific emission lines and continuum isolated by each of the filter passbands (Table~\ref{tbl:summary}). The structures observed in these images will be further discussed in the following sections; here, we briefly summarize the main features of the nebular morphology that is highlighted by the HST/WFC3 image suite.

In all images, the overall SE--NW extension of the nebula, which represents its main fast outflow axis \citep[e.g.,][]{Cox2002}, is apparent. At the shortest (near-UV) wavelengths, the nebula displays the most profound asymmetry along this (SE--NW) direction. Specifically, in both the F243N and F343N images, the regions immediately to the NW of the central star appear brighter than the regions SE of the central star; this asymmetry is a consequence of the smaller dust extinction toward the (forward-facing) NW regions \citep[][]{Montez2018}. The prominent extinction ``hole'' located $\sim$3$''$ NW of the central star discussed in the Chandra X-ray imaging study of \citet[][]{Montez2018}  is especially apparent in both of these near-UV images. The nebula's equatorial dust ``belt,'' as well as a system of dusty filamentary structures located $\sim$5$''$ directly north of the central star, appear as prominent features of the F343N, F487N, F502N images and, to a lesser extent, the F656N and F673N images.  

The near-IR images, covering the range from 1.15 $\mu$m (F110W) to 1.664 $\mu$m (F167N), are less affected by extinction due to intranebular dust, and (as a result) are dominated by the inner elliptical shell and immediately surrounding (``cloverleaf'') structures within the nebula that were previously imaged by HST/NICMOS \citep{Latter2000}. The two images with the highest surface brightness, F502N ($[$O~{\sc iii}$]$) and F656N (H$\alpha$), best reveal the extended system of ring-like dust structures surrounding the nebula that was previously imaged by HST using WFPC2 \citep{Sch2018,Guerrero2020}. 

In all of the WFC3 images (to greater and lesser degrees), jet-like protrusions at position angles (PAs, measured \textbf{counterclockwise} from N) of $\sim$120$^\circ$ (ESE) and $\sim$300$^\circ$ (WNW) are seen superimposed on the aforementioned nebular structures, with the WNW protrusion particularly prominent. Lesser protrusions appear at PAs of $\sim$10$^\circ$ (nearly due N) and $\sim$200$^\circ$ (SSW). In addition, a complex of dusty ejecta is seen oriented nearly due S (most notably in the F343N, F487N, F502N, and F656N images); this south-directed ejecta complex has a faint counterpart at PA $\sim$340$^\circ$ (NNW) that is most apparent in the F502N and F656N images.
\renewcommand{\baselinestretch}{1.0}

\section{Line Ratio Images}\label{sec:analysis}
\subsection{Nebular extinction maps from H recombination lines}\label{sec:Hlines}

\begin{figure}
    \centering
    \includegraphics[width=0.99\textwidth]{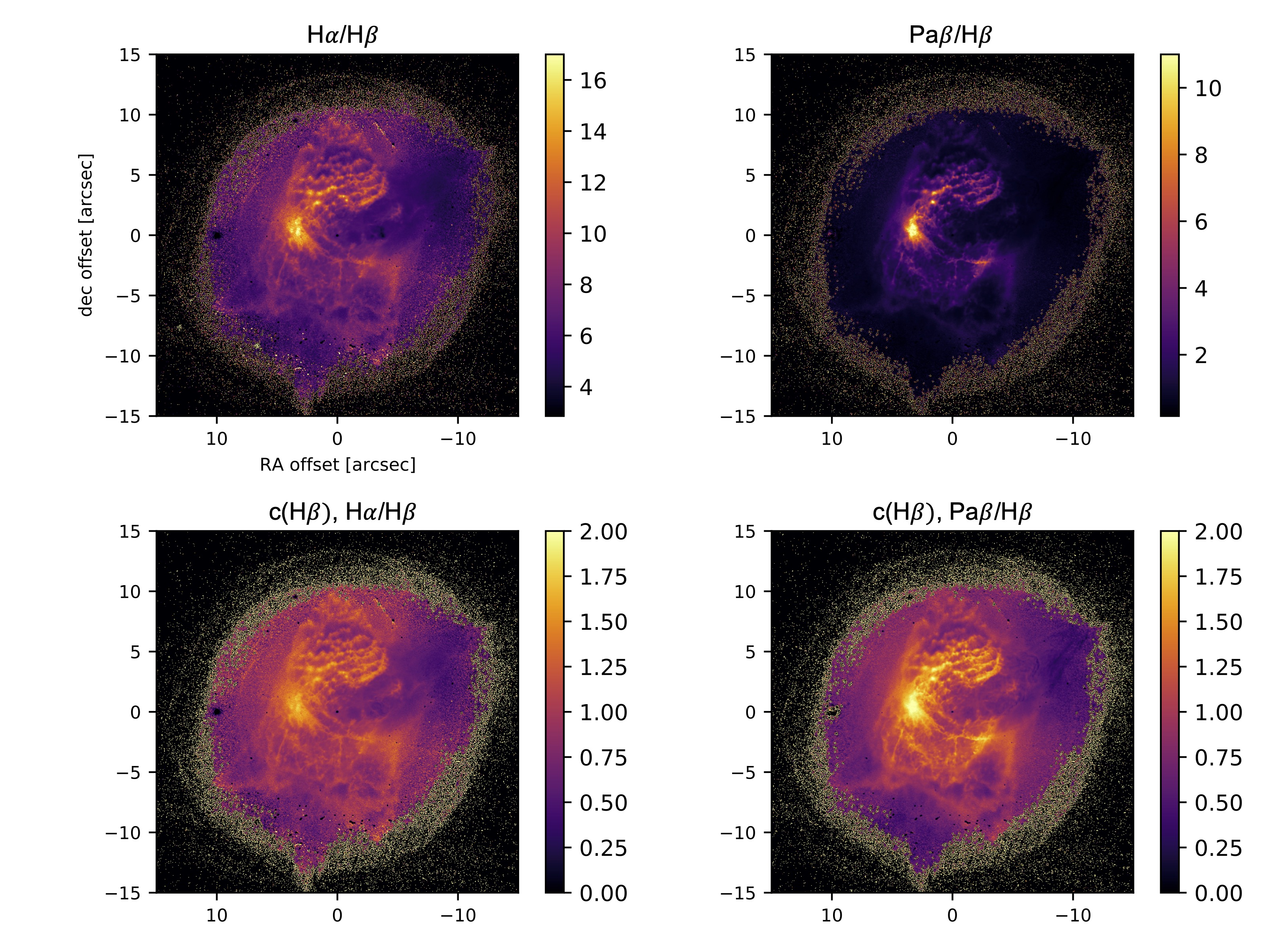}
    \caption{\textit{Top panels:} H$\alpha$/H$\beta$ (left) and Pa$\beta$/H$\beta$ (right) line ratio images, as obtained from the F656N/F487N and (F128N$-$F130N)/F487N image ratios (respectively). The lower display limits are fixed at the theoretically predicted ratios of 2.85 and 0.162, respectively, to highlight regions where extinction is present. \textit{Bottom panels:} c(H$\beta$) calculated from the H$\alpha$/H$\beta$ (left) and Pa$\beta$/H$\beta$ (right) line ratio images.}
    \label{fig:extMap40}
\end{figure}

The three H recombination line images obtained as part of the WFC3 image suite --- i.e., H$\beta$, H$\alpha$, and Pa$\beta$, as obtained through WFC3 filters F487N and F656N and from the F128N$-$F130N difference image, respectively --- provide a means to map dust extinction across the nebula at the $\sim$0.1$''$ resolution of HST/WFC3. To do so, we followed the methodology described in \citet[][]{Kastner2022}; i.e., we constructed H$\alpha$/H$\beta$ and Pa$\beta$/H$\beta$ line ratio images, and compared these per-pixel line ratios with the theoretically predicted ratios, so as to obtain the spatial distribution of the extinction parameter $c(H\beta)$ for a given choice of reddening law. To construct $c(\mathrm{H}\beta)$ images from the H$\alpha$/H$\beta$ and Pa$\beta$/H$\beta$ line ratio images, we adopt intrinsic relative intensities for H$\alpha$/H$\beta$ and Pa$\beta$/H$\beta$ of 2.85 and 0.162, respectively, which correspond to a ``canonical'' values of nebular electron temperature ($10^4$ K) and electron density ($10^4$ cm$^{-3}$), with very little sensitivity to either parameter \citep[e.g.,][]{Osterbrock1989}.
We then adopt a uniform value of $R_V=3.1$ to generate the $c(\mathrm{H}\beta)$ images, with the caveat that $R$ depends on grain size, and grain size would not realistically be uniform across the entire PN. 

The resulting $c(\mathrm{H}\beta)$ images are shown in the bottom panels of Fig. \ref{fig:extMap40}. Ideally, if the $R_{\lambda}$ value used is correct, the c(H$\beta$) images obtained from the two line ratios should be identical. Indeed, the c(H$\beta$) images, shown in the bottom panels of Fig.4, are nearly identical in morphology. The key differences in values occur in the nebula’s waist and the eastern region outlining the north lobe, with the $c(\mathrm{H}\beta)$ map derived using Pa$\beta$/H$\beta$ having systematically larger values than the $c(\mathrm{H}\beta)$ map obtained from H$\alpha$/H$\beta$. Specifically, values across the $c(\mathrm{H}\beta)$ map obtained from Pa$\beta$/H$\beta$ range from $\sim0.9$ (lobes) to $\sim2.1$ (nebula waist), whereas the values in the $c(\mathrm{H}\beta)$ map obtained from H$\alpha$/H$\beta$ lie in the range  $\sim0.9-1.7$. Other, more subtle differences between the extinction maps could be due to the relatively large variation in the PSFs between Pa$\beta$ and H$\beta$, which would affect the resolution of structures across the Pa$\beta$/H$\beta$ ratio image moreso than in the H$\alpha$/H$\beta$ ratio image, and any dust scattering of H$\alpha$ emission that may be present.

Overall, the results for $c(\mathrm{H}\beta)$ are consistent with previous studies that mapped extinction across NGC 7027 \citep[][and references therein]{Kastner2002,Montez2018}. In particular, previous extinction maps also feature extinction maxima along the waist of the nebula, as well as a prominent gap or ``hole'' in extinction located $\sim$5--10$''$ to the WNW of the central star, a region that is also bright in soft X-rays \citep{Montez2018}.
The apparent lack of obscuring nebular material at this location in NGC 7027 is likely due to the disruptive effects of a fast, collimated outflow, as is discussed in detail in \S~\ref{sec:lineDiag}.

However, in contrast to earlier extinction studies, the $\sim$0.1$''$ resolution of the line ratio images and (hence) $c(\mathrm{H}\beta)$ maps obtained from the HST/WFC3 images (Figure~\ref{fig:extMap40}) reveal the dust structures within NGC 7027 in unprecedented detail. North of the CSPN lies dusty lobe structure consisting of a series of ripple-like filaments that terminate at the lobe's northernmost extension, $\sim$10$''$ from the CSPN. In the southern regions of the inner nebula, opposite the CSPN from the northern filament structures, the $c(\mathrm{H}\beta)$ map reveals dust rims and arcs that appear to outline bow-shock-like structures.  These features, along with the large knot along the waist of the nebula, appear more prominent in the $c(\mathrm{H}\beta)$ image derived from Pa$\beta$/H$\beta$ (Figure~\ref{fig:extMap40}, lower right panel). All of the dust structures revealed in the $c(\mathrm{H}\beta)$ maps in Figure~\ref{fig:extMap40} appear to have direct counterparts in near-IR H$_2$ emission \citep[see, e.g., Fig.~3 in][]{Latter2000}, indicating that these structures constitute dense regions of dust and molecular gas that have survived the processes of photoionization and shock-driven ionization that are ongoing in the inner nebula (see next).

In Appendix \ref{sec:extCorr}, we describe how these c(H$\beta$) images are used to perform extinction corrections for several of the HST/WFC3 images.

\subsection{High and low excitation lines: ionization gradients}\label{sec:highlow}

\begin{figure}
    \centering
    \includegraphics[width=0.99\textwidth]{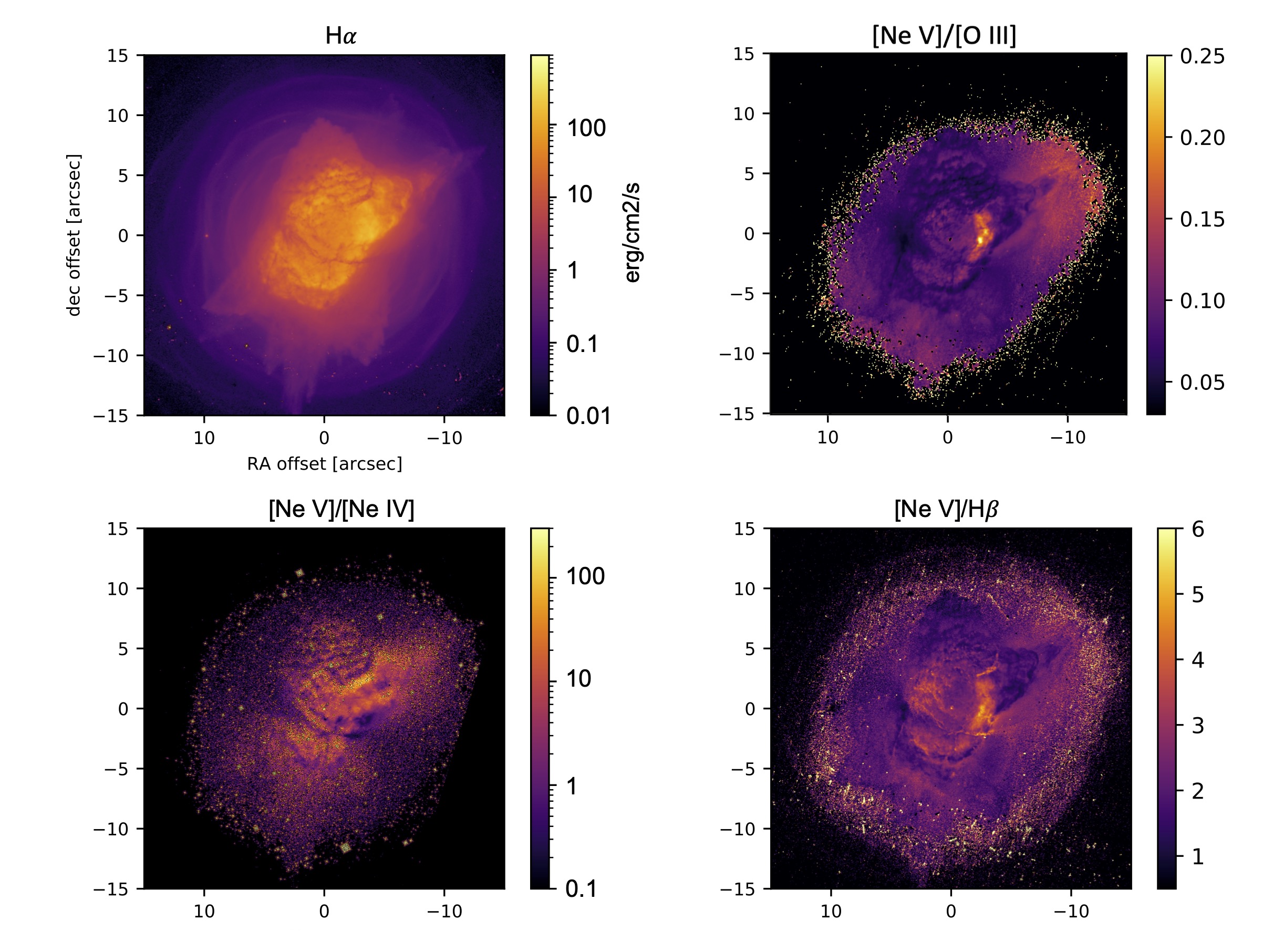}
    \caption{HST/WFC3 line ratio images of NGC 7027 that are potentially diagnostic of photoionization.  \textit{Top right:} $[$Ne{\sc v}$]$/$[$O{\sc iii}$]$ (F343N/F502N); \textit{bottom left:} $[$Ne{\sc v}$]$/$[$Ne{\sc iv}$]$ (F343N/FQ243N); \textit{bottom right:} $[$Ne{\sc v}$]$/H$\beta$ (F343N/F487N). The H$\alpha$ (F656N) image is displayed in the \textit{top left} panel, for reference.}
    \label{fig:ratMap2}
\end{figure}

\begin{figure}
    \centering
    \includegraphics[width=0.99\textwidth]{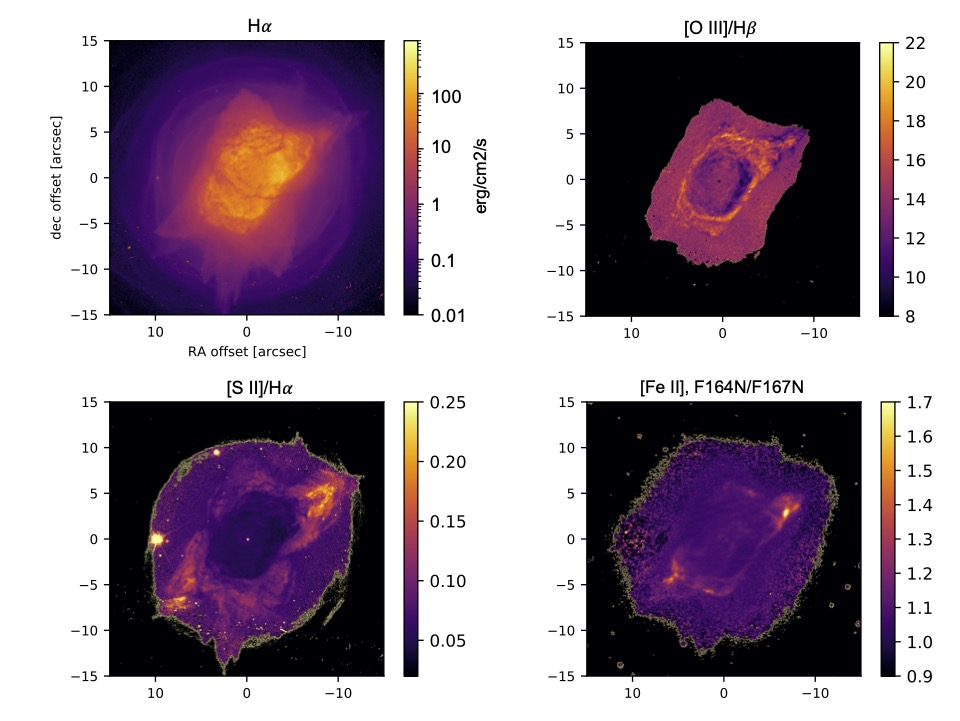}
    \caption{HST/WFC3 line ratio images of NGC 7027 that are potentially diagnostic of excitation by shocks. \textit{Top right:})$[$O{\sc iii}$]$/H$\beta$ (F502N/F487N); \textit{bottom left:} $[$S{\sc ii}$]$/H$\alpha$ (F673N/F656N); \textit{bottom right:} [Fe {\sc ii}]/continuum (F164N/F167N). The H$\alpha$ (F656N) image is displayed in the \textit{top left} panel, for reference.}
    \label{fig:ratMap1}
\end{figure}

In Figs.~\ref{fig:ratMap2} and \ref{fig:ratMap1} we  present a series of emission line and emission line ratio images obtained from the WFC3 image suite. Fig.~\ref{fig:ratMap2} displays the H$\alpha$ emission line and a series of line ratio images obtained for the high-excitation forbidden lines $[$O {\sc iii}$]$, $[$Ne {\sc v}$]$, $[$Ne {\sc iv}$]$, and the H$\alpha$ and H$\beta$ recombination lines. The $[$O {\sc iii}$]$/H$\beta$ ratio line image (Fig.~\ref{fig:ratMap1}) does not require extinction correction because of the small wavelength separation between the two emission lines; this ratio image displays structure essentially identical to that seen in the extinction corrected $[$O {\sc iii}$]$/H$\alpha$ ratio image (as displayed in Fig.~\ref{fig:[OIII]} of Appendix~\ref{sec:extCorr}). 

The $[$O {\sc iii}$]$/H$\beta$ ratio image appears to trace an elliptical shell of efficient O$^+$ ionization (relative to H ionization) with a semimajor axis of $\sim$4$''$ ($\sim$3600 au). This shell marks the zone where photons with energies between 35.1 and 54.9 eV (i.e., those required for ionization of O$^+$ and O$^{++}$, respectively), are plentiful, so as to specifically generate $[$O {\sc iii}$]$ emission.
Closer to the central star (CS), inside of the enhanced $[$O {\sc iii}$]$/H$\beta$ region, the O is likely to be more highly ionized (O$^{+2}$, O$^{+3}$, etc) by photons with energies $>$54.9 eV, while outside the shell of enhanced $[$O {\sc iii}$]$/H$\beta$, most photons with energies $>$35.1 eV have been absorbed by the interior regions of ionized nebular gas.

This inner zone of very high excitation is revealed in the various line ratio images involving $[$Ne~{\sc iv}$]$ and $[$Ne~{\sc v}$]$ in Fig. \ref{fig:ratMap2}. The energies required for ionization of O$^+$ and Ne$^{3+}$ are 35.1 eV and 97.1 eV, respectively, such that one expects the $[$Ne{\sc v}$]$ emitting region to lie interior to that of $[$O~{\sc iii}$]$. That is, photons with energies $>$97.1 eV would be absorbed closer to the central star, via ionization of Ne$^{+3}$, producing $[$Ne {\sc v}$]$ emission; while photons with energies $<$97.1 eV but $\geq$35.1 eV can escape to ionize O$^+$, generating $[$O {\sc iii}$]$ outside of the $[$Ne {\sc v}$]$ emission region. 
This expectation is indeed confirmed by the $[$Ne {\sc v}$]$/H$\beta$ ratio image; this image shows a patchy inner ring of bright $[$Ne~{\sc v}$]$ that resides interior to that of the enhanced $[$O {\sc iii}$]$, which appears as a bright ring in $[$O~{\sc iii}$]$/$[$Ne~{\sc v}$]$. However, the $[$O~{\sc iii}$]$/$[$Ne~{\sc v}$]$ ratio image also reveals prominent, extended $[$O {\sc iii}$]$ emission that overlaps shocked regions, suggesting that photoionization effects could be boosted by shock mechanisms occurring within the nebula (see below).

Fig.~\ref{fig:ratMap1} displays emission line and emission line ratio images featuring H$\alpha$, $[$O {\sc iii}$]$, and the low excitation forbidden lines $[$S {\sc ii}$]$ and $[$Fe {\sc ii}$]$. Extinction correction is not necessary for $[$S {\sc ii}$]$/H$\alpha$ because of the close proximity between $[$S {\sc ii}$]$ (0.676$\mu$m) and H$\alpha$ (0.656$\mu$m) emission lines. These line ratio images are displayed alongside an image of the ratio of $[$Fe~{\sc ii}$]$ 1.64 $\mu$m forbidden line emission with respect to neighboring 1.67 $\mu$m continuum emission (we use this F164N/F167N ratio image in preference to a F164N$-$F167N difference image, due to its superior signal-to-noise ratio). The 1.67~$\mu$m continuum emission is likely due to free-free emission and hence (like H$\alpha$) traces ionized H. As discussed in \cite{Kastner2022}, both the 0.676~$\mu$m $[$S {\sc ii}$]$ and  (especially) 1.64 $\mu$m $[$Fe {\sc ii}$]$ emission lines are potential tracers of shock-driven excitation. Fig.~\ref{fig:ratMap1} reveals that the peaks in $[$S {\sc ii}$]$ and $[$Fe {\sc ii}$]$ emission (with respect to ionized H) in NGC 7027 are highly localized at the tips of the breakout regions to the southeast and northwest of the central star, with fainter, enhanced emission from both lines seen surrounding the bright ring of enhanced $[$O {\sc iii}$]$/H$\alpha$. Furthermore, the brightest $[$Fe~{\sc ii}$]$ lies just interior to the peak in $[$S {\sc ii}$]$/H$\alpha$. As described in \S~\ref{sec:lineDiag}, this nested structure likely traces shock excitation at the outskirts of NGC 7027. Furthermore, the $[$O {\sc iii}$]$/H$\alpha$ ratio image also displays a finger-like protrusion to the NW that closely follows the zones of brightest $[$Fe {\sc ii}$]$ and $[$S~{\sc ii}$]$ emission. This correspondence suggests that the same collimated jet or shock mechanism that generates $[$Fe {\sc ii}$]$ and $[$S {\sc ii}$]$ emission in the NW breakout structure could be contributing to ionization of O$^+$ and, hence, to $[$O {\sc iii}$]$ emission, which would be consistent with studies indicating that [O {\sc iii}] emission can trace shock zones (in addition to photoionization) in PNe \citep{Guerrero2013}.

The $[$Ne {\sc v}$]$/$[$Ne {\sc iv}$]$ ratio image in Fig.~\ref{fig:ratMap2} should also provide a map of the reach of highly ionizing radiation, with the caveat that the wide bandwidth of the F343N [Ne v] filter encompasses several permitted O {\sc iii} lines that are relatively bright in NGC 7027 (see Table~\ref{tab:filters}). However, it is apparent that the signature of intranebular extinction remains strong in this ratio image, despite both component images being subject to the extinction correction procedure described in the Appendix. Specifically, the dark features around the nebula's waist and toward its northern reaches in the $[$Ne {\sc v}$]$/$[$Ne {\sc iv}$]$ image correspond to regions where extinction maps reveal large amounts of dust and dust scattering (see Fig. \ref{fig:extMap40}). Intriguingly, the overall morphology of this image resembles that of the broad-band Chandra X-ray image of NGC 7027 \cite[][see below]{Montez2018}. This suggests that the regions of enhanced $[$Ne {\sc v}$]$ emission (relative to $[$Ne {\sc iv}$]$) might trace a zone of heat conduction from the shocked, X-ray-emitting ($\sim10^6$ K) plasma to the ($\sim10^4$ K) photoionized gas. Alternatively, the $[$Ne {\sc v}$]$/$[$Ne {\sc iv}$]$ image could be affected by scattering, especially in the outer regions of the point-symmetric structures, given that scattering is more efficient at smaller wavelengths, and the scattered UV photons can most easily escape through the low-extinction ``holes'' formed by the blowouts associated with regions of shocks (see below).

\section{Extinction toward the Central Star}\label{sec:CS}

The emission line ratio-based approach to measuring intranebular extinction described in \S~\ref{sec:Hlines} cannot be applied to determine the extinction along the line of sight directly toward the CSPN of NGC 7027, as the star is a bright continuum source. Such an extinction estimate is essential in order to estimate the CSPN's present effective temperature T$_{eff}$ and luminosity $L$, and thereby ascertain the present mass and evolutionary state of the CSPN via comparison with theoretical models \citep[e.g.,][]{Miller2016}. Hence, we estimated the extinction toward the CSPN via comparison of the star's spectral energy distribution with stellar atmosphere models calculated for hot central stars of PNe. To this end, measurements of the CSPN flux, corrected for local nebular background, were made for 9 HST/WFC3 images spanning a wavelength range from $0.3435\ \mu$m to $1.6677\ \mu$m. Fig.~5 displays these observed fluxes. 

\begin{figure}
    \centering
    \includegraphics[width=0.99\textwidth]{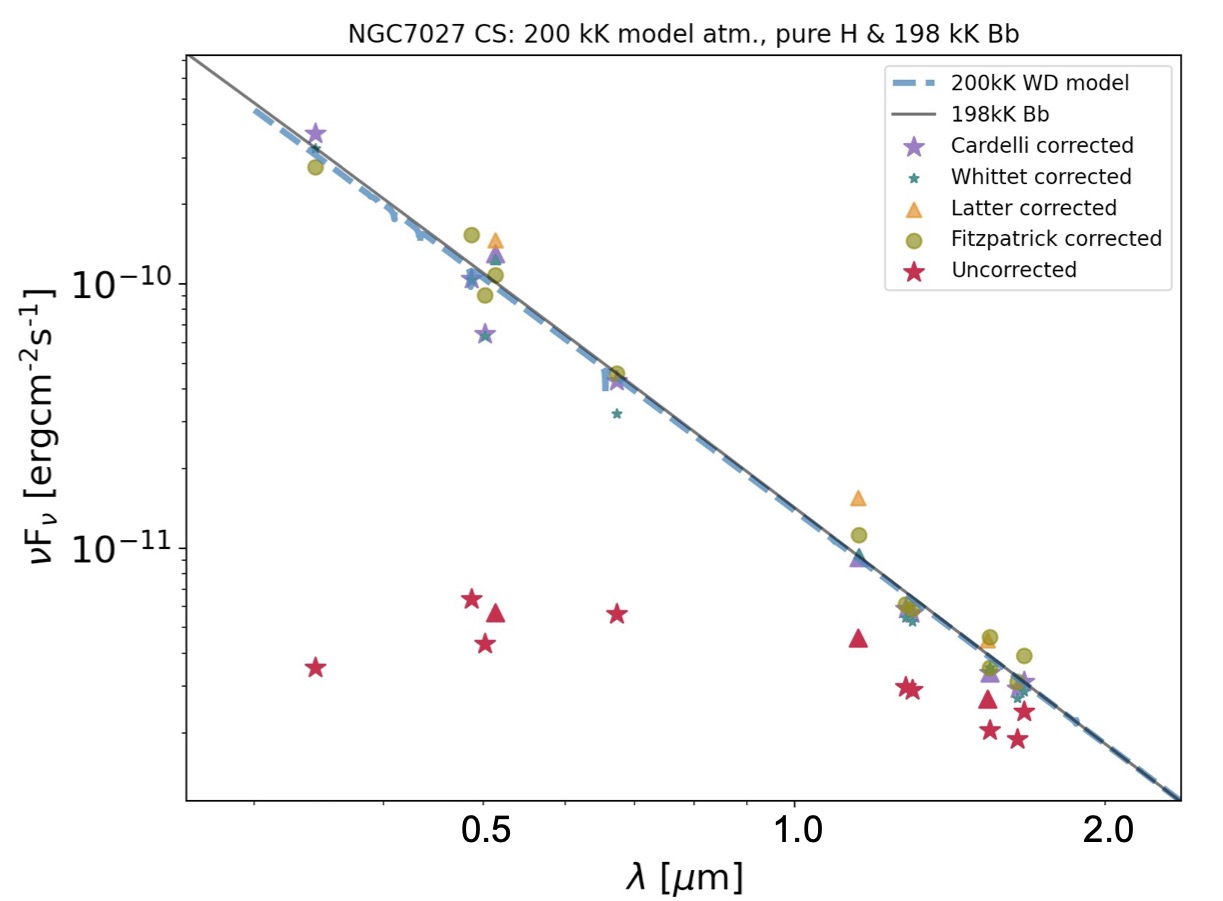}
    \caption{The spectral energy distribution of the NGC 7027 CS compared to a 200 kK pure H WD model atmosphere for $\log (g)=7.0$ (dashed blue line) and 198 kK blackbody (solid gray line) before and after correcting fluxes for extinction. Red stars are uncorrected fluxes obtained from HST/WFC3 image photometry; red triangles are uncorrected fluxes from \citet{Latter2000}. Extinction-corrected fluxes adopting reddening laws from \citet{Cardelli1989}, \citet{Whittet1992}, and \citet{Fitzpatrick2019} are represented as purple, teal, and green symbols; corrected fluxes from \cite{Latter2000} are represented as yellow triangles. }
\end{figure}

The resulting spectral energy distribution (SED) was then compared with a T\"{u}bingen NLTE model atmosphere\footnote{http://astro.uni-tuebingen.de/~rauch/TMAF/TMAF.html} \citep{atmModels} for a model with $T_{eff} =$ 200 kK
\citep[based on the CS effective temperature inferred by][]{Latter2000}. A pure H model with $\log (g)=7.0$ was selected, for simplicity. 
Including near-IR and optical points from \citet{Latter2000}, the extinction is estimated to be $A_V$ = $2.63$ using the \citet{Cardelli1989} reddening law with $R_\lambda$=3.1, $A_V$ = $2.54$ using the \citet{Whittet1992} reddening law, and $A_V$ = $2.90$ using the \citet{Fitzpatrick2019} reddening law. To correct the CS fluxes, we adopted the mean of these extinction estimates, $A_V$ = $2.69$.
In Fig.~5, uncorrected and extinction corrected flux points are overlaid with a 198 kK blackbody, as suggested by \cite{Latter2000}, for comparison. Since the extinction correction results for the CS are consistent with the 198 kK blackbody approximation, the blackbody model was used to approximate the total flux of the CS for purposes of luminosity estimation. The total bolometric luminosity obtained for the CS of NGC 7027 is $\sim6.2\times10^{3}$ L$_\odot$, which agrees well with previous results, all falling within the range $\sim5.5\times10^{3}-2\times10^{4}$ L$_\odot$ \citep{Bein1996, Masson1989}. 

\renewcommand{\baselinestretch}{1.5}
\begin{table}
\begin{center}
\caption{\sc Properties of NGC 7027 CS}
\label{tbl:models}
\footnotesize
\begin{tabular}{ccccc}
\hline
Temperature & Luminosity &  Age$^a$ & Mass$^a$\ \\
(kK) & (L$_\odot$) & (yrs) & (M$_\odot$) \\
\hline
\hline
155$^b$ & $9.9\times10^3$ $^b$ & $\sim100$ & $\sim$0.7 \\
198$^c$ & $6.2\times10^{3}$ & $<1000$ & $\sim$0.7 \\
\hline
\end{tabular}
\end{center}
\renewcommand{\baselinestretch}{1.0}
\tablecomments{a) Estimated using stellar evolution models in \cite{Bertolami2015}. b) Blackbody effective temperature estimate derived in \cite{Schon2005}. c) Blackbody effective temperature estimate derived in \cite{Latter2000}.}
\end{table}

Table \ref{tbl:models} summarizes the estimated age and present-day mass of NGC 7027's CS obtained from the preceding results for luminosity and effective temperature as well as the CS parameters derived by \cite{Schon2005}, based on stellar evolution models presented in \cite{Bertolami2015}. These age and mass estimates are consistent with results in \citet{Latter2000}, who estimated a post-AGB age of 700 yrs and present-day CS mass of $0.7$ M$_\odot$, as well as with estimates of CS mass and post-AGB age obtained from other post-MS evolutionary models. textbf{Our estimated present-day CSPN mass is somewhat discrepant with (larger than) other estimates, however.} For example, stellar evolution models in \citet{Weidmann2020} yield an age of $\sim1000$ years and final mass of $0.6-0.65$ M$_\odot$ for a CS with a luminosity of $6.2\times10^{3}$ erg s$^{-1}$ and an effective temperature of $198$ kK. From the models presented in \citet{Valenzuela2019}, we would estimate a final mass of $0.616-0.706$ M$_\odot$, while using the core mass vs.\ luminosity relation for M $>$ $0.5$ M$_\odot$ stars in \citet{Vass1994}, we estimate a final CS mass of $\sim\ 0.61$ M$_\odot$. Despite inconsistencies in the mentioned results, these estimates all suggest that the CS is descended from a progenitor of mass $\geq$3 $M_\odot$ \citep[e.g.,][]{Miller2016}.



\section{Evidence for Shocks in NGC 7027}

\begin{figure}
    \centering
    \includegraphics[width=0.85\textwidth]{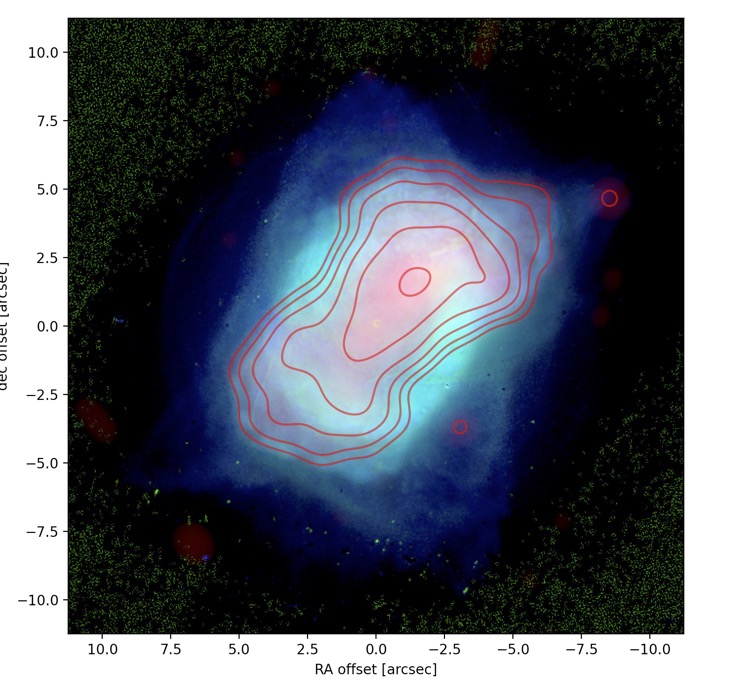}
    \caption{Color composite of Chandra and HST/WFC3 images of NGC 7027 showing overlay of hard X-rays (1.0$-$3.0 keV; red image and contours), extinction corrected $[$Ne {\sc v}$]$ (green image), and extinction corrected H$\alpha$ (blue image). }
    \label{fig:RGBhardX}
\end{figure}

The presence of strong, wind-collision-generated shocks in NGC 7027 was revealed by the Chandra X-ray Observatory detection of extended, luminous X-ray emission from the nebula extending over the  energy range $\sim$0.3-3.0 keV \citep{Kastner2001,Montez2018}.
\cite{Montez2018} estimate that the velocities of the outflows that generate the X-ray-emitting regions of NGC 7027 may be as high as $v_{\mathrm{wind}}\sim500$ km s$^{-1}$, on the basis of the inferred characteristic temperature of the X-ray emitting plasma ($\sim3\times10^6$ K). 
In Fig. \ref{fig:RGBhardX} we present an overlay of the hard (1.0$-$3.0 keV) Chandra X-ray image from \citet{Montez2018} on the HST/WFC3 $[$Ne {\sc v}$]$ and H$\alpha$ images. It is apparent from this Figure that the hard X-ray emission closely traces the edge of the inner, highly ionized ([Ne~{\sc v}]-bright) region of the nebula, with the exception of the equatorial region to the ENE and WSW of the central star. These regions correspond to directions of the largest dust extinction, as seen in Fig. \ref{fig:extMap40}, indicating that the X-ray emission in these regions is likely being heavily absorbed by cooler, surrounding, CNO-rich gas \citep[see discussion in][]{Montez2018}.
Furthermore, the X-ray surface brightness protrusions to the NW and SE of the central star closely correspond to regions of high-velocity gas detected in near-IR H {\sc i} (Br$\gamma$) emission \citep{Cox2002}. 

In this section, we present additional evidence for shocks in NGC 7027, through line ratio diagnostics obtained from the [O {\sc iii}]/H$\alpha$ and [S {\sc ii}]/H$\alpha$ ratio images along with a comparison of nebula cross sections in these line ratios and those of [Fe {\sc ii}] and X-ray surface brightness.
\renewcommand{\baselinestretch}{1.0}

\subsection{Line ratio diagnostics}\label{sec:lineDiag}

Excitation diagrams obtained from diagnostic line ratios, long used to classify AGN \citep[e.g.,][]{Kewley2006}, can, in principle, also be used to distinguish between photoionized regions and fast low-ionization structures (LISs), or shocked material, in PNe \citep[][]{Danehkar2018,Montoro2022}. Here, we follow the methodology described in \cite{Danehkar2018} to quantitatively identify the shocked regions in NGC 7027.

\begin{figure}
    \centering
    \includegraphics[width=1\textwidth]{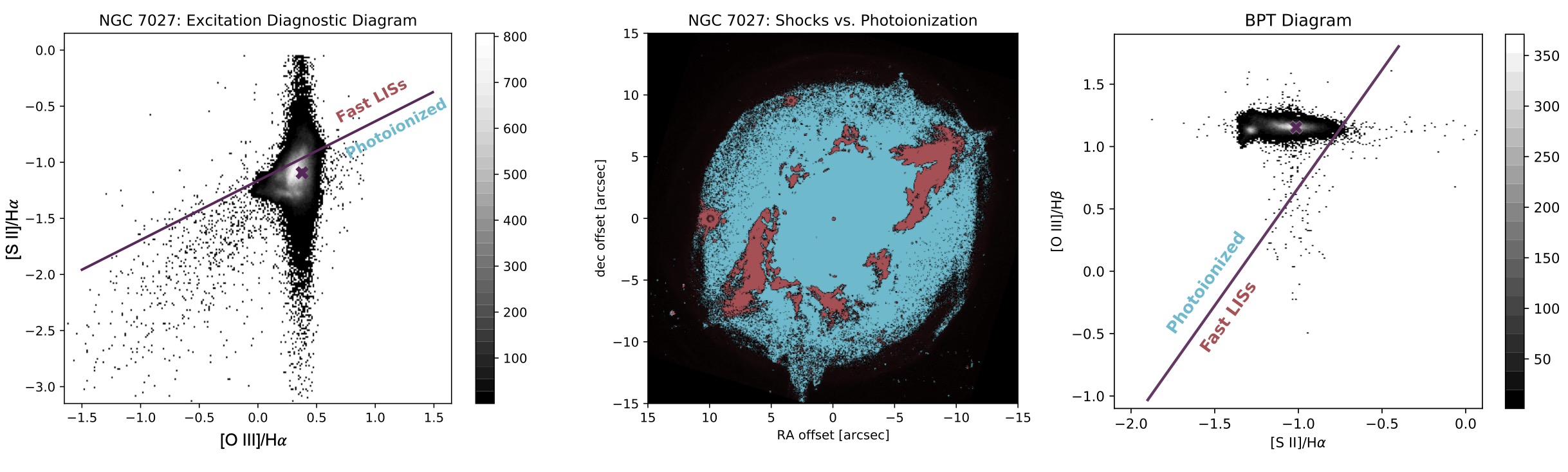}
    \caption{Diagnostic diagram analysis to distinguish between photoionized and shocked regions within NGC 7027. The \textit{left panel} presents an excitation diagram that differentiates between $[$O {\sc iii}$]$ and $[$S {\sc ii}$]$ emission due to fast LISs vs.\ photoionization. The color bar indicates the number of pixels in a given area of the diagram. The mean flux for both line ratios is indicated by the `x'. The nebular photon-shock dividing line, adapted from \cite{Danehkar2018}, is $1.89\log([\mathrm{S\ {\sc II}}]/\mathrm{H}\alpha)+2.0=\log([\mathrm{O\ {\sc III}}]/\mathrm{H}\alpha)$. In the \textit{middle panel}, the H$\alpha$ image of NGC 7027 is mapped with blue and red pixels representing photoionization and fast LISs, respectively, as obtained from the diagram in the left panel. The \textit{right panel} displays a `Baldwin, Phillips, and Terlevich' (BPT) diagram, commonly used to classify AGN \citep{Kewley2006}, using $[$S {\sc ii}$]$/H$\alpha$ and $[$O {\sc iii}$]$/H$\beta$ emission line ratio images. A different nebular photon-shock dividing line, here adapted from \cite{Danehkar2018} as $1.89\log([\mathrm{S\ {\sc II}}]/\mathrm{H}\alpha)+2.46=\log([\mathrm{O\ {\sc III}}]/\mathrm{H}\beta)$, is plotted to differentiate between emission due to fast LISs vs.\ photoionization as well. A similar color bar indicates the number of pixels in a region of the diagram, and an `x' was placed to mark the mean flux of both emission line ratio images.}
    \label{fig:ExDiag}
\end{figure}

A histogram was produced for $\sim6\times 10^5$ pixels in the [O {\sc iii}]/H$\alpha$ and [S {\sc ii}]/H$\alpha$ line ratio images for an area that covers $30'' \times 30''$. The results are presented in the left panel of Fig. \ref{fig:ExDiag}. The dividing line overlaid on the histogram should distinguish between pixels that belong to photoionized regions and fast LISs, wherein the photoionized regions are characterized by enhanced [O {\sc iii}]/H$\alpha$ and suppressed [S {\sc ii}]/H$\alpha$ and fast LISs are characterized by suppressed [O {\sc iii}]/H$\alpha$ and enhanced [S {\sc ii}]/H$\alpha$. \cite{Danehkar2018} defines this nebular photon-shock dividing line to be parallel to a classification line that differentiates between Seyfert and low-ionization nuclear emission-line region (LINER) galaxies. The definition was modified to take into account diagnostic diagrams for axisymmetric simulations of fast LISs and observations of photoionized gas in \citep{Raga2008}.

The excitation diagram was then used to classify each pixel and then to color code an H-alpha intensity image of the nebula. The result is shown in the middle panel of Fig. \ref{fig:ExDiag}, where blue represents photoionized regions and red represents fast LISs. The results indicate that the elliptical interior region of NGC 7027 is dominated by photoionization, while the most extensive shocked regions, according to this analysis, correspond to the same directions as the X-ray protrusions \citep{Montez2018} and the fast flows seen in Br$\gamma$ emission \cite{Cox2002}.

Most PN studies have, by necessity, employed spatially integrated line ratios to determine where nebulae lie on a given excitation diagnostic diagram such as that in the left panel of Fig. \ref{fig:ExDiag} \citep[e.g.,][]{Frew2010,Raga2008}. To illustrate how this analysis would play out for NGC 7027, we have indicated the mean line ratios obtained from the NGC 7027 images, $[$O {\sc iii}$]$/H$\alpha=2.4\pm0.7$, $[$O {\sc iii}$]$/H$\beta=14.3\pm1.4$ and $[$S {\sc ii}$]/$H$\alpha=0.09\pm0.05$, in the diagnostic diagrams--left and right panels in Fig. \ref{fig:ExDiag}.  Note that these mean ratios are within the ``photoionized'' zone of the diagram, as a consequence of the relatively small surface area of the nebula that displays LIS-like ratios. Thus, if only spatially integrated line ratios were available, one might conclude there were no shocks present in NGC 7027. We caution, however, that this young and rapidly evolving PN may represent an unusual case study in the application of spatially resolved line ratio diagnostics to distinguish between shocked and photoionized regions  \citep[see discussion in][]{Montoro2022}.

\newpage
\subsection{Nested shock structure}\label{nest}

\begin{figure}
    \centering
    \includegraphics[width=0.95\textwidth]{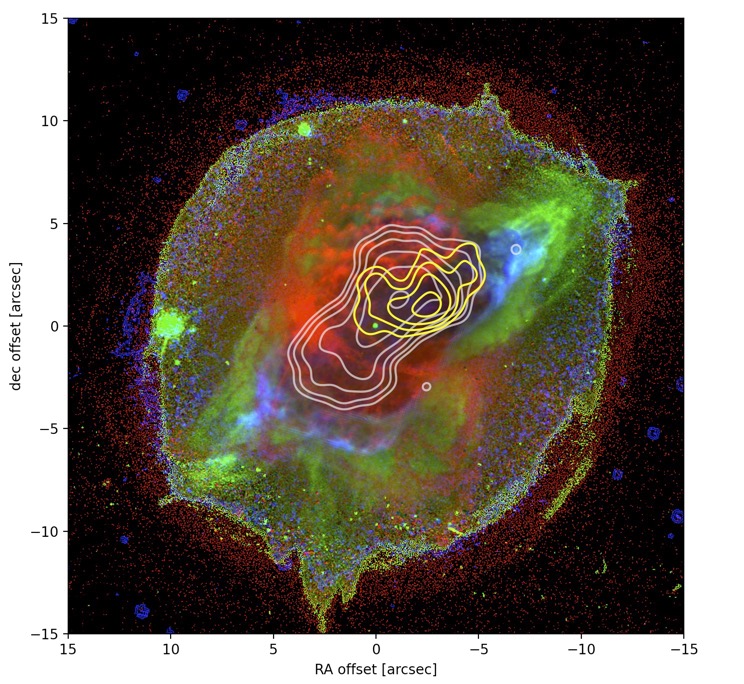}
    \caption{Three-color composite of NGC 7027. Red is F656N/F487N (H$\alpha$/H$\beta$), green is F673N/F656N ($[$S{\sc ii}$]$/H$\alpha$), and blue is F164N/F167N ($[$Fe{\sc ii}$]$). White and yellow contours trace hard and soft X-ray emission, respectively.}
    \label{fig:rgb}
\end{figure}

In Fig. \ref{fig:rgb}, we present a line ratio image overlay for NGC 7027 that compares the positions of bright $[$S {\sc ii}$]$ and $[$Fe {\sc ii}$]$ emission with the X-ray emission morphology as well as the zones of high extinction (as traced by the H$\alpha$/H$\beta$ line ratio). The Figure reveals a stratified structure, wherein the $[$S {\sc ii}$]$ and $[$Fe {\sc ii}$]$ emitting regions display bow shock structures extending well beyond the X-ray emission, and the $[$Fe {\sc ii}$]$  emission nestles nicely inside the $[$S {\sc ii}$]$ emission. This set of nested bright features in X-rays, $[$Fe {\sc ii}$]$, and $[$S {\sc ii}$]$ along the (SW--NE) direction --- which also constitutes the primary direction of fast LISs identified via the diagnostic diagram analysis (Fig.~\ref{fig:ExDiag}) --- serves as clear evidence that the shocks due to fast outflows impinging on the nebular material are strongest in this particular direction.

\begin{figure}
    \centering
    \includegraphics[width=0.95\textwidth]{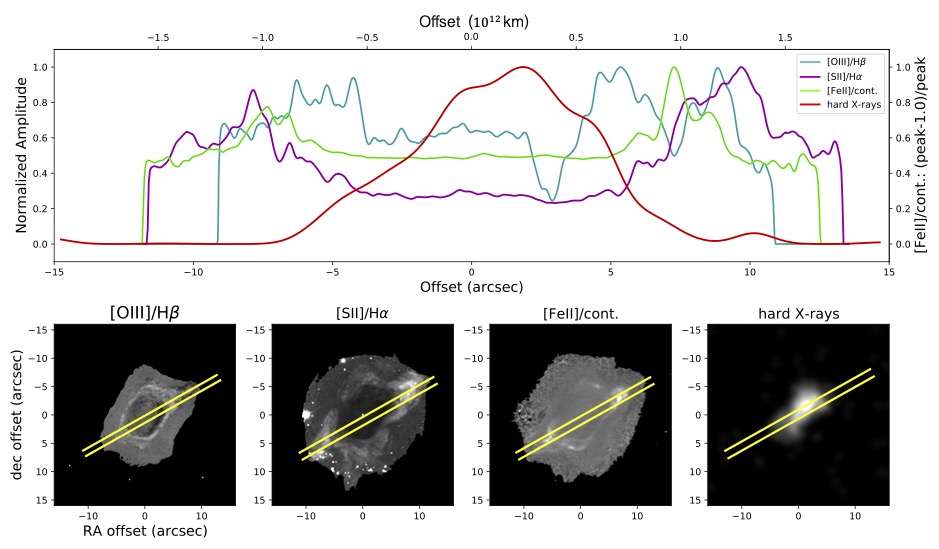}
    \caption{Overlay of line ratio and emission cross sections through NGC 7027. Bottom panels show the images ( [O {\sc iii}]/H$\beta$, [S {\sc ii}]/H$\alpha$, [Fe {\sc ii}], and [energy here] X-rays) from which the cross sections were extracted for the nebula, with the cross sections indicated by yellow lines. The top panel plots the normalized amplitude for each image's cross section vs. the offset from the central star in arcsec (bottom axis) and cm (top axis).}
    \label{fig:cross}
\end{figure}

To further elucidate this structure, we present in Fig.~\ref{fig:cross} a cross-section of the nebula in X-rays, $[$S {\sc ii}$]$, $[$Fe {\sc ii}$]$ as well as $[$O {\sc iii}$]$ emission along this same direction. This cross-section plot confirms the nested structure of these shock tracers, and reveals the detailed structure of the particularly strong shocks that are propagating along this direction within the nebula. As discussed below (\S 7), the clarity of these shocks compared to the nebula's other outflows indicates that the jets emanating along the SW--NE direction could dominate the future shaping process of NGC 7027, perhaps eventually forming a bilobed, pinch-waisted PN that appears as a classical ``butterfly'' nebula, like NGC 6302 \citep[][]{Kastner2022}. 

The nested spatial distributions of the $[$S {\sc ii}$]$ and $[$Fe {\sc ii}$]$  emission lines also should aid in constraining the densities within these regions of NGC 7027. In a study of the symbiotic system BI Crucis, whose excitation is largely collisional, \citet[][and references therein]{Contini2009}  point out that $[$Fe~{\sc ii}$]$ emission should dominate over Fe~{\sc ii} permitted lines if the gas density of a nebula falls between 10$^2 -$10$^4$ cm$^{-3}$, and both $[$Fe~{\sc ii}$]$ and Fe~{\sc ii} lines should be excited if the gas density lies between 10$^6 -$10$^8$ cm$^{-3}$. Similarly, the presence of $[$S~{\sc ii}$]$ emission constrains the gas density to be $<$10$^4$ cm$^{-3}$ \citep{Acker1995}. Since NGC 7027's optical spectrum displays fairly weak permitted Fe~{\sc ii} and S~{\sc ii} lines compared to their forbidden emission counterparts \citep{Zhang2005}, we infer a nebular density of $<$10$^4$ cm$^{-3}$ in the NW and SE lobe structures.  However, these constraints also suggest that the gas from which $[$Fe {\sc ii}$]$ emission arises should have a higher density than regions where $[$S~{\sc ii}$]$ emission is observed, since regions of $[$S {\sc ii}$]$ emission are collisionally de-excited (hence quenched) at a density $<$10$^4$ cm$^{-3}$, but regions of $[$Fe {\sc ii}$]$ emission can reach densities of $\sim$10$^5$ cm$^{-3}$ without having visible Fe {\sc ii} lines. The expectation that $[$Fe {\sc ii}$]$ should trace denser gas than $[$S~{\sc ii}$]$ is indeed consistent with the nested emission structure apparent in Fig.~\ref{fig:cross}.

\section{The shaping of NGC 7027}

Previous sections of this paper have developed the deepest-ever  look into the structure and current state of NGC 7027 and its central star.  In this section, we use these data to attempt to peek back into the source and evolution of mass injection and nebular shaping.

\begin{figure}
    \centering
    \includegraphics[width=0.95\textwidth]{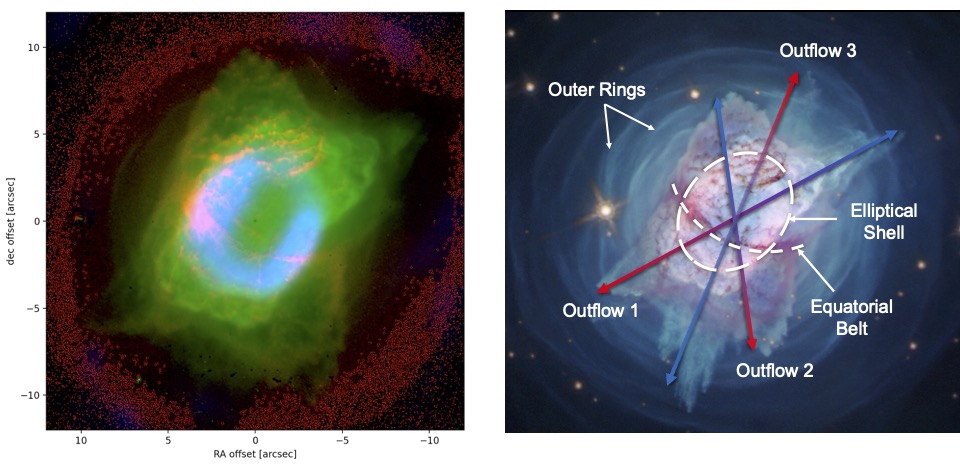}
    \caption{\textit{Left panel}, three color composite of NGC 7027 generated from Pa$\beta$/H$\beta$ (red), extinction corrected H$\alpha$ (green), and 1.3 mm radio continuum \citep[blue; from][]{Bublitz2022ngc7027}. 
    \textit{Right panel}, schematic diagram   
    outlining the structural components discussed in \S~\ref{sec:shaping}, overlaid on a color composite constructed from the WFC3 image suite by STScI staff. 
    Three pairs of jets, labeled as Outflows 1, 2, and 3 \citep[see][]{Cox2002}, are traced with arrows that are color-coded to indicate the relative redshift/blueshift for each outflow component. The nebula's outer ring system, central elliptical shell (long-dashed ellipse), and equatorial belt (short-dashed arc) are also indicated.}
    \label{fig:schem}
\end{figure}

\subsection{A series of shaping events}\label{sec:shaping}

Fig.~\ref{fig:schem} provides an overview of the key structures observed in NGC 7027. The left panel presents an overlay of HST/WFC3 extinction and H$\alpha$ images, as well as a 1.3 mm radio continuum image obtained with the NOEMA interferometer \citep[from][]{Bublitz2022ngc7027}; a schematic diagram of NGC 7027 outlining the primary structural features is presented  in the right panel, overlaid on a color composite constructed from the WFC3 image suite by STScI staff\footnote{https://www.nasa.gov/feature/goddard/2020/hubble-provides-holistic-view-of-stars-gone-haywire}. In the schematic diagram, each of the three primary previously identified outflows \citep[][]{Cox2002,Nakashima2010,Bublitz2022ngc7027} is traced by arrows, with color-coded red and blue extremes to signify their relative redshifts and blueshifts, respectively. The elliptical shell seen in the WFC3 NIR images and radio continuum image is traced by a dashed ellipse. The outer ring system, which is most clearly visible in the H$\alpha$ and [O~{\sc iii}] images (Fig. \ref{fig:rawImgs}), as well as the equatorial belt of NGC 7027 are also indicated in the schematic.

We propose that this complex set of structures is most likely the result of a series of PN shaping events resulting from binary system interactions involving NGC 7027's AGB star progenitor and an unseen compact companion. Binary interactions while the primary star was still on the AGB likely resulted in the spirals seen surrounding NGC 7027 (in scattered light) in the $[$O {\sc iii}$]$ and H$\alpha$ images \citep[at least 9 have been identified;][]{Corradi2004}. Such ring or spiral patterns in AGB ejecta are potential signatures of the presence of a wide (separation $\sim$1--10 au) binary system 
\citep[e.g.,][and references therein]{MastrodemosMorris1999,Corradi2004,Chen2020} and, indeed, have been detected in the ejected envelopes of AGB stars that are suspected binary systems \citep[e.g., R Scl;][and references therein]{Maercker2012}.
The estimated post-AGB age of NGC 7027's central star, $<1$ kyr (\S~\ref{sec:CS}), is significantly less than the estimated age of the rings, $\sim2$ kyr \citep{Su2004,Guerrero2020}, supporting the idea that these outer rings formed before the CS left the AGB. The nebula's prominent equatorial belt might then trace the binary's orbital plane, where the bulk of the AGB star's ejected envelope mass would have accumulated \citep[e.g.,][]{MastrodemosMorris1999}.

Of the three main outflows noted in Fig. \ref{fig:schem}, Outflow 1 (the focus of the discussion in \S~\ref{nest}) displays the most pronounced evidence for strong shocks, at the present epoch. The fact that Outflows 2 and 3 appear less prominent or absent in the shock tracers [S {\sc ii}] and [Fe {\sc ii}] relative to Outflow 1 may indicate that Outflows 2 and 3 are older than (were ejected prior to) Outflow 1, despite their similar projected extents. Indeed, Outflows 2 and 3 could have occurred during NGC 7027's post-AGB but pre-planetary nebula (PPN) phase \citep{SahaiTrauger1998,Sahai2000,Akashi2021}, perhaps during repeated close binary encounters that culminated in the unveiling of the central star (see below). These two pairs of jets have triangular shapes that may be bow shocks generated by narrow, collimated jets. Furthermore, Outflow 3 displays a striking misalignment relative to Outflows 1 and 2, wherein Outflow 3 appears to be oriented close to NGC 7027's equatorial plane \citep[][and references therein]{Bublitz2022ngc7027}. Some PPNe display a similar combination of recent, collimated outflows --- some with similarly surprising misalignments --- impinging on older, more slowly expanding ring/spiral structures  \citep[AFGL 2688 perhaps being the best-studied example of this phenomenon; e.g.,][]{Cox2000,Balick2012}.  

The central elliptical shell of NGC 7027 (Fig.~\ref{fig:schem}) likely consists of the ejected material from the former AGB star that has been swept up by a fast, quasi-spherical post-AGB wind. The age of the CS since leaving the AGB has been estimated consistently to be $\sim$100-1000 yrs (\S~\ref{sec:CS}). However, since this post-AGB CS age pertains to a single star rather than a central star subject to binary system interactions that would, presumably, truncate AGB evolution, this estimate is most likely an upper limit. Given that the youngest outflow, Outflow 1, may have been ejected as recently as $\sim$100 yrs ago \citep{Cox2002}, it is possible that this outflow and the unveiling of the CS occurred essentially simultaneously. A thin shell of [Fe~{\sc ii}] emission is seen surrounding the elliptical shell (Fig.~\ref{fig:schem}), connecting the shell to the shocked regions of Outflow 1 (as seen in [S~{\sc ii}] and [Fe~{\sc ii}]), perhaps providing evidence for this simultaneity.  

\subsection{Did NGC 7027 undergo an ILOT?}

The foregoing sequence of events in the shaping history of NGC 7027 appears consistent with a model of formation for bipolar PNe formation via Intermediate-Luminosity Optical Transient (ILOT) ejection events \citep{Soker2012}. This (ILOT) model offers an explanation for the formation of bi-lobed PNe that display evidence for multiple, paired mass ejections. \cite{Soker2012} set out the following criteria to identify PNe that may have been formed by ILOT events: the PN (1) contains structures or components that exhibit a linear relationship between velocity and distance; 
(2) is bipolar in structure,
indicating the presence of a binary system with orbital separation of $\sim1$ AU; (3) has lobe expansion velocities exceeding $\sim100$ km s$^{-1}$; and (4) has a total fast outflow kinetic energy in the range $\sim10^{46}$--$10^{49}$ erg. \cite{Soker2012} further point to NGC 6302 as an exemplar of a bipolar PN that was generated by an ILOT event some $\sim$2000 yr ago \citep[see also][]{Kastner2022}.

Indeed, NGC 7027 would appear to satisfy all four ILOT criteria. With regard to criteria 1--3: its cloverleaf-shaped H$_2$ emitting region exhibits a velocity field proportional to radius \citep{Cox2002}; the nebula displays (nascent) bipolar structure, in the form of its dusty equatorial belt and collimated outflows (especially the youngest, Outflow 1; Fig.~\ref{fig:schem}); and the presence of [Fe {\sc ii}] and X-ray emission, as well as high-velocity Br$\gamma$ emission \citep{Cox2002}, along Outflow 1 indicates outflow speeds in excess of $\sim100$ km s$^{-1}$, at least along this (SE--NW) direction (see \S~\ref{nest}). NGC 7027 also appears to meet the fourth (energy) criterion: adopting the mass of $0.1$ M$_{\odot}$ estimated by \citet{Santander2013} for their ``high-velocity molecular blobs'', and an assumed velocity of 100 km s$^{-1}$, we arrive at a total energy of $\sim2\times10^{46}$ erg for its collimated outflows. With a post-AGB age of just $\sim$100--1000 yr (\S \ref{sec:CS}) and at least one collimated outflow (Outflow 1) that is of similar dynamical age \citep{Cox2002}, NGC 7027 may therefore be the youngest known example of an ILOT-generated bipolar PN.



\section{Summary}

The young, rapidly evolving, nearby NGC 7027 is among the best objects for studying PN shaping and excitation mechanisms. We have obtained and analyzed a comprehensive suite of narrow-band HST/WFC3 images for this nebula, spanning a range from near-UV ($\sim$245 nm) to near-IR ($\sim$1665 nm). Our analysis and main results can be summarized as follows.

\begin{itemize}
    \item H$\alpha$/H$\beta$ and Pa$\beta$/H$\beta$ emission line ratio images are constructed and a per-pixel comparison is made with the theoretically predicted ratios, so as to obtain the spatial distribution of the extinction parameter c(H$\beta$). 
    Overall, the results for c(H$\beta$) are consistent with previous studies that mapped extinction across NGC 7027, featuring extinction maxima along the waist of the nebula, as well as a prominent gap in extinction located $\sim5''-10''$ to the WNW of the central star; however, the extinction maps reveal the dusty substructures within NGC 7027 in unprecedented detail. The c(H$\beta$) images are used to perform extinction corrections for several of the HST/WFC3 images (see Appendix \ref{sec:extCorr}).
    \item Aperture photometry performed on the WFC3 images and near-IR and optical measurements from \citet{Latter2000} are used to construct the spectral energy distribution of the CSPN from the near-UV to near-IR. Analysis of these data yields an estimate of extinction of A$_V = 2.69$ toward the central star.
    The resulting CSPN luminosity is $6.2\times10^{3}$ erg s$^{-1}$ for an assumed effective temperature of $198$ kK \citep{Latter2000}. These parameters indicate that the central star is descended from a progenitor of $\geq$3 $M_\odot$, has a present-day mass of $\sim$0.6--0.7 $M_\odot$, and left the AGB within the last $\sim$1000 years.
    \item Emission line ratio images constructed for [O {\sc iii}]/[Ne {\sc v}], [Ne {\sc v}]/[Ne {\sc iv}], [Ne {\sc v}]/H$\beta$, [O {\sc iii}]/H$\beta$, [S {\sc ii}]/H$\alpha$, and [Fe {\sc ii}]/continuum reveal the detailed ionization structure of the nebula (\S~4). This structure appears best explained as a juxtaposition of photoionized and shock-ionized zones within the nebula.
    An excitation diagram is used 
    to assess the spatial regimes of photoionized vs.\ shocked nebular material (\S~\ref{sec:lineDiag}). This analysis reveals that the outflow structures oriented SW--NE --- Outflow 1 in Fig.~\ref{fig:schem} --- display the strongest shocks. This conclusion is bolstered by the consistent orientations of the extended X-ray, [S {\sc ii}], and [Fe {\sc ii}] emitting regions along this same direction. The nested spatial distributions of these tracers (and [O{\sc iii}]) reveal the detailed shock structure (and constrain the gas densities) along Outflow 1.
    \item A timeline for the shaping of NGC 7027 is proposed, in which a close binary system has produced the nebula's ring system and equatorial belt, and subsequent binary interactions, possibly in a series of short-lived eruptions (the ILOT scenario), then generate the multipolar jets observed. This model awaits confirmation through a study of the proper motions of the complex NGC 7027 jet system, via these WFC3 and archival WFPC2 HST images, so as to establish more precise dynamical ages.
\end{itemize}

\section*{Acknowledgements}
\begin{acknowledgements}
We thank Jesus Taola and the anonymous referee for helpful comments that improved the clarity of this paper. Based on observations made with the NASA/ESA Hubble Space Telescope, obtained at the Space Telescope Science Institute (STScI), operated by the Association of Universities for Research in Astronomy, Inc., under NASA contract NAS5-26555. These observations are associated with program \#15953. Support for this program was provided by NASA through STScI grant HST-GO-15953.001-A to RIT. All of the HST/WFC3 data presented in this paper were obtained from the Mikulski Archive for Space Telescopes (MAST) at the Space Telescope Science Institute. The specific observations analyzed can be accessed via \dataset[DOI]{10.17909/hm6d-qh67}.
\end{acknowledgements}

\bibliography{references}

\appendix
\section{Image extinction correction}\label{sec:extCorr}

The $c(\mathrm{H}\beta)$ images obtained from the H recombination line images (\S \ref{sec:Hlines}) in principle provide a means to correct the extinction from the HST/WFC3 suite images. This correction was performed, initially, on $[$Ne {\sc v}$]$ and $[$O {\sc iii}$]$ images using $c(\mathrm{H}\beta)$ images as a starting point 
for extinction calculations at 0.343 $\mu$m ($[$Ne {\sc v}$]$) and 0.501 $\mu$m ($[$O {\sc iii}$]$). The relation between $c(H\beta)$ and $E(B-V)$ is described by \cite{1983Howarth} as $c(\lambda)=0.4(R+0.53)E(B-V)$. For $R_{\lambda}=3.1$, this relation becomes $c(\lambda)=1.45E(B-V)$. Using the reddening law $f(\lambda)$ from \cite{Cardelli1989}, the extinction at a given wavelength is then obtained as $A_{\lambda}=f(\lambda) R_{\lambda}E(B-V)$. 

\begin{figure}
    \centering
    \includegraphics[width=0.99\textwidth]{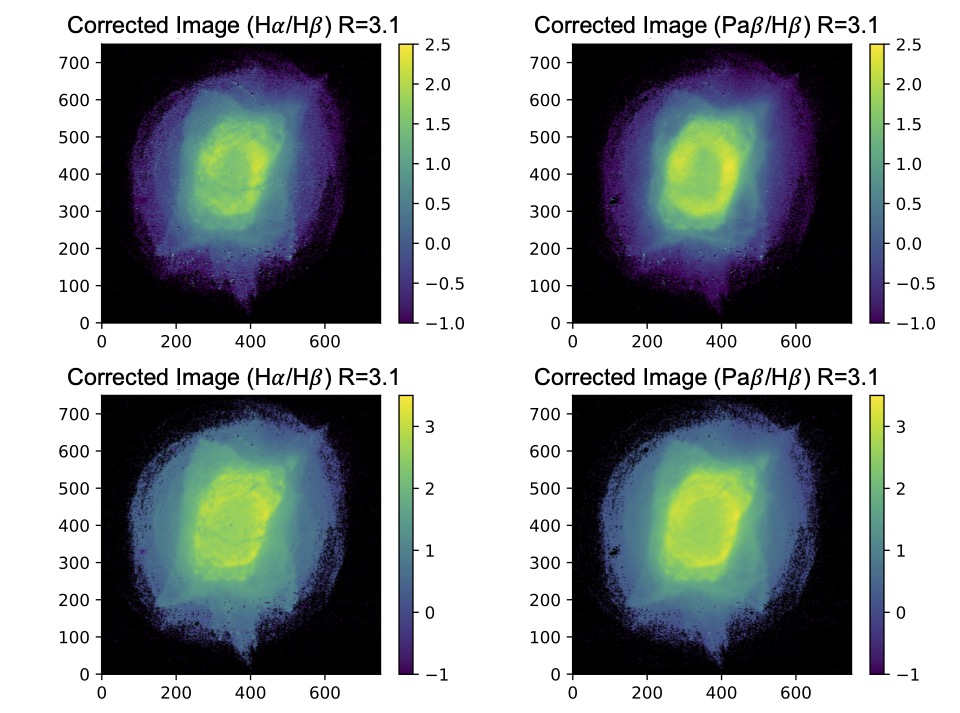}
    \caption{Comparison of extinction-corrected $[$Ne~{\sc v}$]$  (top panels) and $[$O {\sc iii}$]$ (bottom panels) images as obtained using $c(\mathrm{H}\beta)$ images derived from H$\alpha$/H$\beta$ (left panels) and Pa$\beta$/H$\beta$ (right panels), for $R_{\lambda}=3.1$. Images are presented in their native rotation and units (counts s$^{-1}$, on a log intensity scale), and the axes correspond to pixel size.}
    \label{fig:NeVcorr}
\end{figure}

The images in Fig. \ref{fig:NeVcorr} show results of the extinction corrections performed for $[$Ne~{\sc v}$]$ and $[$O~{\sc iii}$]$, respectively. The $c(\mathrm{H}\beta)$ images obtained from Pa$\beta$/H$\beta$ and H$\alpha$/H$\beta$ yield slight differences in extinction calculation for both $[$Ne {\sc v}$]$ and $[$O {\sc iii}$]$. For $[$Ne {\sc v}$]$, the $c(\mathrm{H}\beta)$ image derived from H$\alpha$/H$\beta$ produced traces of slight overcorrection within the nebula's left dusty region and a possible undercorrection near the nebula's waist. This becomes clearer when comparing results generated using the $c(\mathrm{H}\beta)$ image derived from the Pa$\beta$/H$\beta$ image. Undercorrection in the nebula's waist caused by the H$\alpha$/H$\beta$ image can also be identified in $[$O {\sc iii}$]$ results as remnants of a dust lane. 

\begin{figure}
    \centering
    \includegraphics[width=0.7\textwidth]{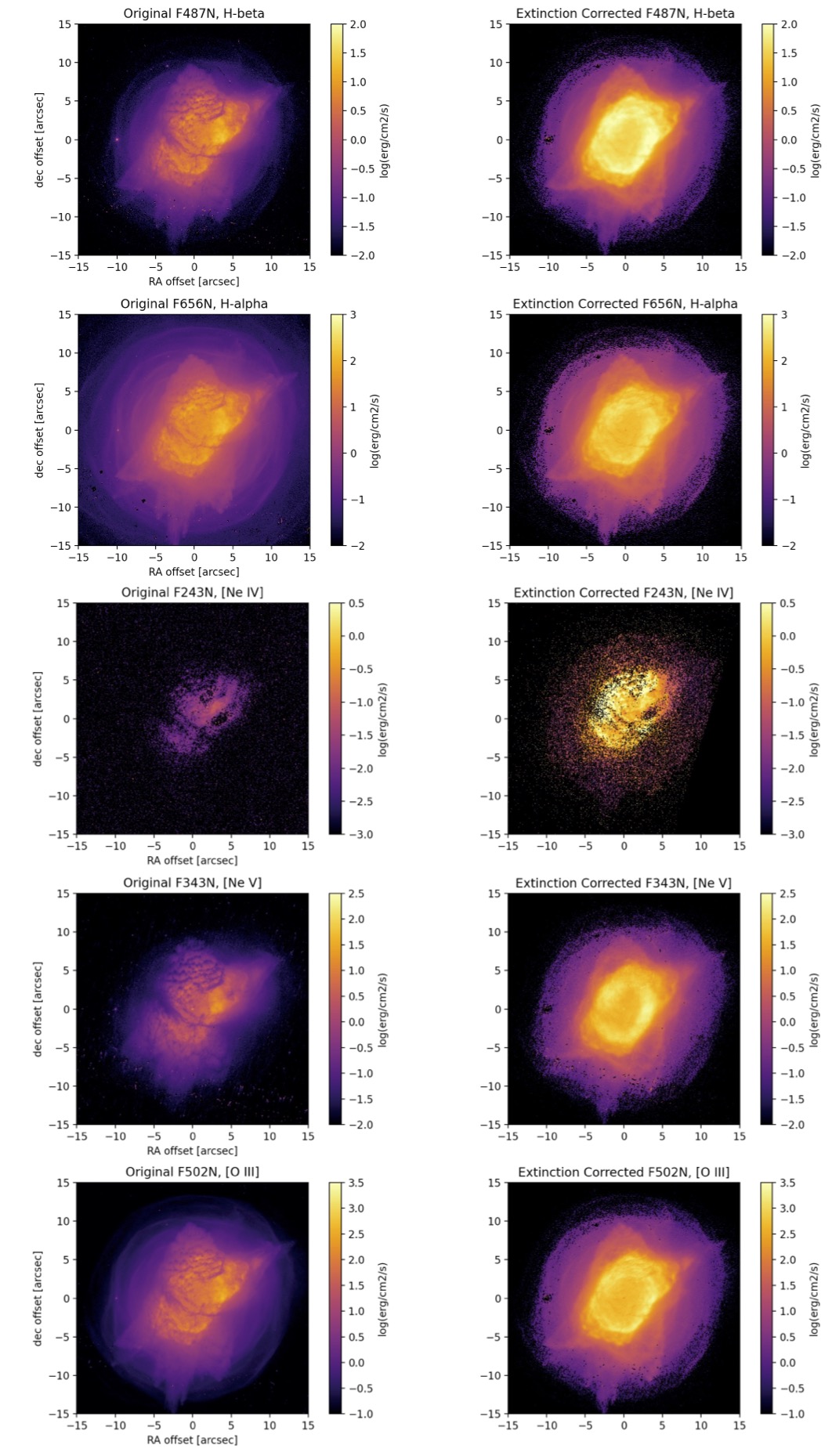}
    \caption{Comparison of original H$\beta$, H$\alpha$, $[$Ne {\sc iv}$]$, $[$Ne {\sc v}$]$, and $[$O {\sc iii}$]$ images (left panels) with the extinction-corrected images (right panels).}
    \label{fig:AllCorr}
\end{figure}

Extinction corrections for $[$Ne {\sc v}$]$, $[$Ne {\sc iv}$]$, H$\alpha$, H$\beta$, and $[$O {\sc iii}$]$ images were performed using $c(\mathrm{H}\beta)$ images derived from the Pa$\beta$/H$\beta$ line ratio map. Fig. \ref{fig:AllCorr} compares the extinction-corrected images with their raw HST/WFC3 counterparts.
In all but the $[$Ne {\sc iv}$]$ image, the extinction corrections recover the full extent of the bright elliptical shell seen in the (dust-penetrating) WFC3 near-IR images. The correction is least successful for the $[$Ne {\sc iv}$]$ image, likely because of the poor signal-to-noise ratio.

\begin{figure}
    \centering
    \includegraphics[width=0.99\textwidth]{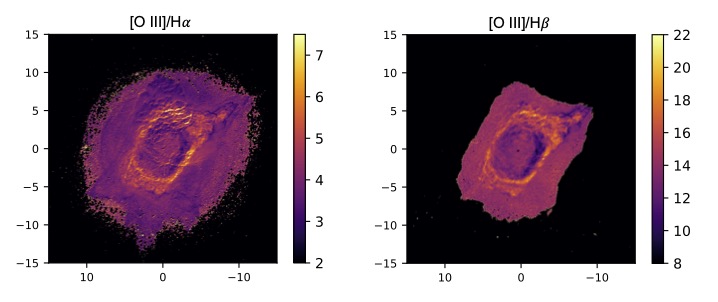}
    \caption{Comparison of extinction corrected $[$O{\sc iii}$]$/H$\alpha$ (\textit{left}),and $[$O{\sc iii}$]$/H$\beta$ (\textit{right}) line ratio images. }
    \label{fig:[OIII]}
\end{figure}

Fig. \ref{fig:[OIII]} compares the $[$O{\sc iii}$]$/H$\beta$ line ratio image with the extinction corrected $[$O{\sc iii}$]$/H$\alpha$ line ratio image. A few differences between these images --- primarily dust lane surrounding the nebula and ripples in the N lobe --- are apparent. Overall, however, the morphologies are very similar, verifying that the extinction corrections are reasonable. 
The extinction correction process was also performed on $[$Ne {\sc iv}$]$, H$\alpha$, and H$\beta$ for line ratio images in Fig. \ref{fig:ratMap2}.

\end{document}